\documentclass[aps,prd,twocolumn,
comment, 
superscriptaddress,groupedaddress,
amsmath,amssymb,floatfix,nofootinbib]
{revtex4}  
\usepackage{float}
\usepackage{graphicx}
\usepackage{bm}
\usepackage[usenames,dvipsnames]{color}
\usepackage{slashed}
\usepackage{hyperref}
\usepackage{slashed}

\begin{document}

%

\let\a=\alpha      \let\b=\beta       \let\c=\chi        \let\d=\delta
\let\e=\varepsilon \let\f=\varphi     \let\g=\gamma      \let\h=\eta
\let\k=\kappa      \let\l=\lambda     \let\m=\mu
\let\o=\omega      \let\r=\varrho     \let\s=\sigma
\let\t=\tau        \let\th=\vartheta  \let\y=\upsilon    \let\x=\xi
\let\z=\zeta       \let\io=\iota      \let\vp=\varpi     \let\ro=\rho
\let\ph=\phi       \let\ep=\epsilon   \let\te=\theta
\let\n=\nu
\let\D=\Delta   \let\F=\Phi    \let\G=\Gamma  \let\L=\Lambda
\let\O=\Omega   \let\P=\Pi     \let\Ps=\Psi   \let\Si=\Sigma
\let\Th=\Theta  \let\X=\Xi     \let\Y=\Upsilon
%


\def\cA{{\cal A}}                \def\cB{{\cal B}}
\def\cC{{\cal C}}                \def\cD{{\cal D}}
\def\cE{{\cal E}}                \def\cF{{\cal F}}
\def\cG{{\cal G}}                \def\cH{{\cal H}}
\def\cI{{\cal I}}                \def\cJ{{\cal J}}
\def\cK{{\cal K}}                \def\cL{{\cal L}}
\def\cM{{\cal M}}                \def\cN{{\cal N}}
\def\cO{{\cal O}}                \def\cP{{\cal P}}
\def\cQ{{\cal Q}}                \def\cR{{\cal R}}
\def\cS{{\cal S}}                \def\cT{{\cal T}}
\def\cU{{\cal U}}                \def\cV{{\cal V}}
\def\cW{{\cal W}}                \def\cX{{\cal X}}
\def\cY{{\cal Y}}                \def\cZ{{\cal Z}}


\def\be{\begin{equation}}
\def\ee{\end{equation}}
\def\bea{\begin{eqnarray}}
\def\eea{\end{eqnarray}}
\def\bm{\begin{matrix}}
\def\em{\end{matrix}}
\def\bpm{\begin{pmatrix}}
    \def\epm{\end{pmatrix}}

{\newcommand{\lsim}{\mbox{\raisebox{-.6ex}{~$\stackrel{<}{\sim}$~}}}
{\newcommand{\gsim}{\mbox{\raisebox{-.6ex}{~$\stackrel{>}{\sim}$~}}}
\def\mpl{M_{\rm {Pl}}}
\def\gev{{\rm \,Ge\kern-0.125em V}}
\def\tev{{\rm \,Te\kern-0.125em V}}
\def\mev{{\rm \,Me\kern-0.125em V}}
\def\ev{\,{\rm eV}}

\title{\boldmath   The $eejj$ Excess Signal at the LHC and Constraints on Leptogenesis}
\author{Mansi Dhuria}
\email{mansi@prl.res.in}
\affiliation{Physical Research Laboratory, Navrangpura, Ahmedabad 380 009, India}
\author{Chandan Hati}
\email{chandan@prl.res.in} 
\affiliation{Physical Research Laboratory, Navrangpura, Ahmedabad 380 009, India}
\affiliation{Indian Institute of Technology Gandhinagar, Chandkheda, Ahmedabad 382 424, India.}
\author{Raghavan Rangarajan}
\email{raghavan@prl.res.in}
\affiliation{Physical Research Laboratory, Navrangpura, Ahmedabad 380 009, India}
\author{Utpal Sarkar}
\email{utpal@prl.res.in} 
\affiliation{Physical Research Laboratory, Navrangpura, Ahmedabad 380 009, India}

\begin{abstract}
	
	We review the  non-supersymmetric (Extended) Left-Right Symmetric Models (LRSM) and low energy $E_6$-based models to investigate if they can explain both the recently detected excess $eejj$ signal at CMS and leptogenesis. The $eejj$ excess can be explained from the decay of the right-handed gauge bosons ($W_R$) with mass $\sim \tev$ in certain variants of the LRSM (with $g_{L}\neq g_{R}$). However such scenarios can not accommodate high-scale leptogenesis. Other attempts have been made to explain leptogenesis while  keeping the $W_{R}$ mass almost within the reach of the LHC by considering the resonant leptogenesis scenario in the context of the LRSM for relatively large Yukawa couplings. However this may not be feasible due to washout of the lepton asymmetry by certain processes. Therefore we consider three effective low energy subgroups of the superstring inspired $E_{6}$ model having a number of additional exotic fermions which provides a rich phenomenology to be explored. We however find that these three effective low energy subgroups of $E_6$ too cannot explain both the $eejj$ excess signal and leptogenesis simultaneously.
	
\end{abstract}
\maketitle


\section{Introduction}
\label{Introduction}
  One of the most popular extensions of the Standard Model (SM) of particle physics is the Left-Right Symmetric Model (LRSM) \cite{Pati:1974yy}. The weak interactions of the LRSM are governed by the gauge group $SU(2)_{L} \times SU(2)_{R} \times U(1)_{B-L}$  where $B-L$ is the difference between baryon and lepton numbers. In such a model, the right-handed gauge bosons $(W_R)$ decay in a manner very similar to their left-handed counterparts except that the left-handed neutrino ($\nu_L$) gets replaced by its right-handed counterpart $N_R$. Now $N_R$ may be a Dirac particle, decaying to a ``proper-sign" lepton or a  Majorana particle which can decay to either sign lepton. It  can further decay via a virtual $W_{R}$ emission or via mixing with $\nu_L$ giving a two lepton two jet signal.

 The CMS Collaboration at the LHC at CERN has announced their results for the $W_{R}$ search at a center of mass energy of $\sqrt{s}=8 \tev$ and $19.7 \rm{fb}^{-1}$ of integrated luminosity. Using the cuts $p_{T}> 60\gev, \vert \eta \vert <2.5 (p_{T}> 40\gev, \vert \eta \vert <2.5)$ for the leading (subleading) electron and selecting events with $m_{ee}>200 \gev$ a total of $14$ events were observed in the energy bin $1.8 \tev< M_{eejj}<2.2 \tev$ compared to $4$ events expected from the SM background giving a $2.8\sigma$ local excess in the $pp\rightarrow ee+2j$ channel  \cite{Khachatryan:2014dka}. The excess of $eejj$ events has been explained to be due to $W_R$ decay by embedding the LRSM  in a class of $SO(10)$ model in Ref. \cite{Deppisch:2014qpa}  and by considering general flavour mixing in the LRSM in Ref. \cite{Saavedra2014}.  Additional tests to study right-handed currents at LHC are proposed in Ref. \cite{Heikinheimo-2014}.

  However confirmation of these excess events for the given  range of the $W_{R}$ mass has severe implications for the leptogenesis mechanism  \cite{Fukugita:1986hr}, which offers a very attractive possibility to explain the baryon asymmetry of the universe.  The seesaw mechanism \cite{Minkowski:1977sc} which provides a natural solution to the smallness of neutrino masses, offers a  mechanism for generating a lepton asymmetry (and hence a $B-L$ asymmetry) before the electroweak phase transition, which then gets converted to the baryon asymmetry of the universe via  $B+L$ violating anomalous processes  in equilibrium \cite{Fukugita:1986hr, Kuzmin:1985mm}. The lepton asymmetry can be generated in two possible ways. One way is via the decay of right-handed Majorana neutrinos ($N$) which does not conserve lepton number \cite{Fukugita:1986hr}; another way is via the decay of very heavy Higgs triplet scalars with lepton number violating interactions \cite{Ma:1998dx}. In the conventional LRSM, the right-handed neutrinos interact with the $SU(2)_R$ gauge bosons. By taking into account the effect of the scattering processes involving such interactions of $W_{R}$ on the primordial lepton asymmetry of the universe,  phenomenologically successful high-scale leptogenesis requires $M_{W_R}$ to be very heavy for both the cases $M_{N}> M_{W_R}$ and $M_{W_R}> M_{N}$ \cite{Ma:1998sq}. Thus, an observed $1.8 \tev < M_{W_{R}}<2.2 \tev $ implies that the decay of right-handed neutrinos can not generate the required lepton asymmetry  of the universe. Furthermore, since the $W_{R}$ interactions erase any primordial $B-L$ asymmetry, the observed baryon asymmetry of the universe must be generated at a scale lower than the $SU(2)_{R}$ breaking scale. Attempts have been made to explain the required amount of lepton asymmetry in the context of resonant leptogenesis \cite{Flanz:1994yx} while pushing the mass of $W_R$ to as low as $3\tev$ for relatively large Yukawa couplings {\cite{Frere:2008ct, Dev:2014iva}}. However in Ref. \cite{Dhuria:inprep} we have found that the presence of certain lepton number violating processes involving the doubly charged  right-handed Higgs triplet in the LRSM  which stay in equilibrium close to the electroweak  phase transition for $M_{W_{R}}$ in the range of a few $\tev$ will result in washing out of any lepton asymmetry created above the electroweak  phase transition. Thus, the $W_R$ mass is required to be quite high compared to the CMS signal range to have a successful resonant leptogenesis scenario. Similar arguments hold true  even for the extended LRSM models which can be formed  by extending the gauge group of the LRSM with additional $U(1)$'s. Therefore we have considered generalized non-supersymmetric LRSM variants motivated by the low-energy subgroups of superstring inspired $E_{6}$ theories. These models are particularly interesting because in addition to having a gauge structure similar to the conventional LRSM, they also have a number of additional exotic fermions, thus providing a rich phenomenology to be explored. To this end, we examine these models to explore if the CMS excess signal can be explained while simultaneously allowing high-scale leptogenesis to generate the observed baryon asymmetry of the universe.  
 
 The outline of the article is as follows. In section {\bf \ref{ELRSM}}, we first discuss the particle content and  $B-L$ breaking scale of the Left-Right Symmetric Model. Then we argue that the $B-L$ breaking scale will be lower than the $SU(2)_R$ breaking  scale even in the extended LRSM. We then discuss how the CMS signal (if it is indeed due to $W_{R}$ decay) rules out the possibility of high-scale leptogenesis and also mention certain lepton number violating processes involving the doubly charged  right-handed Higgs triplet in (Extended) LRSM  which stay in equilibrium close to the electroweak  phase transition. These can rule out the possibility of $\tev$-scale resonant leptogenesis with the $W_R$ mass in the few $\tev$ range.  In section {\bf \ref{E6}}, we first discuss the phenomenology of low energy subgroups of $E_6$ group.  Then we show that though one of the subgroups allows high-scale leptogenesis, there does not exist any  effective low energy subgroups of $E_6$  which can explain both the CMS $eejj$  excess as well as leptogenesis \footnote{Since we consider non-supersymmetric LRSM, we assume that supersymmetry gets broken at a very high-scale in the low energy effective subgroups of $E_6$ and therefore supersymmetric particles are not really relevant for our analysis.}. In section {\bf \ref{conc}}, we conclude with our results.
\section{(Extended) Left Right Symmetric Model (LRSM) and constraints from leptogenesis }
\label{ELRSM}
In the LRSM  the leptons and the quarks transform under the gauge group $SU(2)_{L} \times SU(2)_{R} \times U(1)_{B-L}$  as
\bea{\label{2.1}}
l_{L}&&=\left(\bm \nu  \\ e \em\right)_{L} : (2,1,-1), \ \    l_{R}=\left(\bm N  \\ e \em \right)_{R} : (1,2,-1), \nonumber\\
Q_{L}&&=\left(\bm u  \\ d \em\right)_{L} : (2,1,\frac{1}{3}), \ \    Q_{R}=\left(\bm u  \\ d \em\right)_{R} : (1,2,\frac{1}{3}).
\eea
The Higgs sector of the LRSM consists of one bi-doublet $\Phi$ and two triplet $\Delta_{L,R}$ complex scalar fields with the transformations
\bea{\label{2.2}}
\Phi &=&\bpm \Phi^{0}_{1} & \Phi^{+}_{1} \\ \Phi^{-}_{2} & \Phi^{0}_{2}\epm : (2,2,0),\nonumber\\
\Delta_{L} &=& \bpm \frac{\Delta^{+}_L}{\sqrt{2}} & \Delta^{++}_L \\ \Delta^{0}_L & -\frac{\Delta^{+}_L}{\sqrt{2}}\epm_{L} : (3,1,2),\nonumber\\
 \Delta_{R}&= &\bpm \frac{\Delta^{+}_R}{\sqrt{2}} & \Delta^{++}_R \\ \Delta^{0}_R & -\frac{\Delta^{+}_R}{\sqrt{2}}\epm_{R} : (1,3,2).
\eea
The left-right symmetry can be spontaneously broken to reproduce the Standard Model and the smallness of the neutrino masses can be taken care of by the see-saw mechanism. The symmetry breaking mechanism follows the scheme
\bea{\label{2.3}}
&& SU(3)_{C}\times SU(2)_{L}\times SU(2)_{R}\times U(1)_{B-L}\nonumber\\
 &&\rightarrow SU(3)_{C}\times SU(2)_{L}\times U(1)_{Y}\nonumber\\
&&\rightarrow SU(3)_{C} \times U(1)_{EM}
\eea
Being aware of the above we now  turn the table around and ask the question that if the CMS signal is indeed due to the decay of $W_{R}$ corresponding to  $SU(2)_{R}$ breaking then can we conclusively say that (one of) the $U(1)$(s) in the left-right symmetric scheme (and its $U(1)$ extensions) is necessarily $U(1)_{B-L}$. If so then the next question is at what scale does it get broken. We start with an arbitrary $U(1)$ (where we do not identify the $U(1)$ charge with $B-L$) in the LRSM gauge group and then generalize the scheme to include more than one $U(1)$. Consider the scheme $SU(2)_{L}\times SU(2)_{R}\times U(1)_{X}$, where the charge of the quark doublet under $U(1)$ is assumed to be $X_{Q}$ and that of the lepton pair is assumed to be $X_{l}$. Under $U(1)_{X}$ the fields transform as 
\bea {\label{2.4}}
l_{L} &:& (2, 1, X_{l}), l_{R} : (1, 2, X_{l}),\nonumber\\
 Q_{L} &:& (2, 1, X_{Q}), Q_{R} : (1, 2, X_{Q}),\nonumber\\
\Phi &:&(2, 2, 0), \Delta_{L} : (3, 1, -2X_{l}), \Delta_{R} : (1,3, -2X_{l}).
\eea
Now we consider a scenario where in the first stage the right-handed triplet $\Delta_{R}$ acquires a Vacuum Expectation Value (VEV)
\be {\label{2.5}}
\langle \Delta_{R}\rangle=\frac{1}{\sqrt 2}\bpm 0 & 0 \\ v_{R} & 0 \epm
\ee
 which breaks the $SU(2)_{R}$ symmetry to give the right-handed neutrino a Majorana mass and to produce massive $W^{\pm}_{R}$, $Z_{R}$ bosons. The next stage involves breaking the electroweak symmetry at some lower energy where the bi-doublet Higgs and left-handed Higgs triplet get VEVs giving mass to $W^{\pm}_{L}$ and $Z_{L}$ gauge bosons \footnote{ Note that giving VEV to $\Delta_{L}$ is not necessary, however if such a scheme is allowed then left-handed fermions get both Majorana and Dirac masses.}. It turns out that in such a scheme $X$ can only be $B-L$  and the combination $\tau^{3}_{L}+\tau^{3}_{R}+\frac{1}{2} {\bf{1}}_{B-L}$ is the only unbroken generator satisfying the modified Gell- Mann-Nishijima formula
  \be {\label{2.8}}
 Q = T_{3L}+T_{3R}+\frac{B-L}{2}.
 \ee
 The $B-L$ symmetry can be violated either simultaneously or at a scale below the $SU(2)_{R}$ breaking scale.\\
 
 Next we consider the Extended LRSM such as $SU(3)_{C}\times SU(2)_{L}\times SU(2)_{R}\times U(1)_{X}\times U(1)_{Z}$, where we do not identify either $X$ or $Z$ with $B-L$. Then also we can argue that the $B-L$ breaking scale is lower than or equal to the $SU(2)_{R}$ breaking scale. The argument goes as follows. We perform an $SO(2)$ rotation on the gauge fields $(A^{X}, A^{Z})$ and choose a new basis $U(1)^{\prime}_{X} \times U(1)^{\prime}_{Z}$ such that the charge of $\Phi$ for one of the two groups, say $U(1)^{\prime}_{X}$, is zero. At this point we identify $U(1)^{\prime}_{X}$ with $B-L$. So the transformations of the Higgs fields are given by
  \be {\label{2.9}}
 \Phi : (2, 2, 0, Q^{\Phi}_{Z}),  \Delta_{L} : (3, 1, 2, Q^{\Delta_{L}}_{Z}),  \Delta_{R} : (1,3, 2, Q^{\Delta_{R}}_{Z}).
 \ee
 So this reduces to the standard LRSM scenario if the additional $U(1)^{\prime}_{Z}$ breaks at a scale higher than the $SU(2)_{R}$ breaking scale. This chain of arguments continue for any arbitrary number of $U(1)$ extensions of the LRSM. Thus, $B-L$ gets broken either simultaneously with the $SU(2)_{R}$ or else at a scale lower than the $SU(2)_{R}$ breaking scale in the LRSM or any extension of the LRSM with arbitrary numbers of $U(1)$'s.
 
  The most stringent constraints on the $W_{R}$ mass for  successful high-scale leptogenesis come from the $SU(2)_{R}$ interactions \cite{Ma:1998sq}. To have successful leptogenesis in the case $M_{N}>M_{W_{R}}$ the condition that the process
 \be
 e_{R}^{-}+W_{R}^{+}\rightarrow N_{R} \rightarrow e_{R}^{+}+W_{R}^{-}
 \ee
 goes out of equilibrium translates into the condition 
 \be
 M_{N}\gsim 10^{16} \gev
 \ee
  with $m_{W_{R}}/m_{N}\gtrsim 0.1$. Now for the case $M_{W_{R}}>M_{N} $ leptogenesis can happen either at $T\simeq M_{N}$ or at $T>M_{W_{R}}$ but at less than  $B-L$ breaking scale. For $T\simeq M_{N}$, the condition that 
 the scattering processes that maintain the equilibrium number density for $N_R$ go out of equilibrium
 reduces to
 \be
  M_{W_{R}}\gtrsim 2\times 10^{5} \gev (M_{N}/10^{2} \gev)^{3/4}.
  \ee
  For leptogenesis at $T>M_{W_{R}}$ the most relevant scattering process is
  \be
   W_{R}^{\pm}+W_{R}^{\pm}\rightarrow e_{R}^{\pm} + e_{R}^{\pm}
  \ee
 through $N_{R}$ exchange and the condition for this process to go out of equilibrium gives
 \be
 M_{W_{R}}\gtrsim 3\times 10^{6} \gev (M_{N}/10^{2} \gev)^{2/3}.
 \ee
  Thus it follows that if the CMS excess is indeed a $W_{R}$ signal with the mass of the $W_{R}$ in the range $1.8 \tev < M_{W_{R}}<2.2 \tev $ then for hierarchical neutrino masses ($M_{N_{3R}}\gg M_{N_{2R}}\gg M_{N_{1R}}=m_{N}$) it is not possible to generate the required baryon asymmetry of the universe from high-scale leptogenesis.
 
The possibility of  generating the required lepton asymmetry with a considerably low value of the $W_R$ mass has been discussed in the context of the resonant leptogenesis scenario \cite{Flanz:1994yx}. It has also been pointed out that successful low-scale leptogenesis with a quasi-degenerate right-handed neutrinos mass spectrum requires an absolute lower bound of 18 TeV on the $W_R$ mass \cite{Frere:2008ct}. Recently it was shown that just the right amount of  lepton asymmetry can be produced even for a substantially  low value of the $W_R$ mass ($M_{W_{R}}>3 \tev$) \cite{Dev:2014iva} by considering relatively large Yukawa couplings. However there are certain lepton number violating processes which are ignored in the aforementioned analysis. In particular, below the left-right symmetry breaking scale, the lepton number violating scattering processes $e_{R}^{\pm}W_{R}^{\mp}\rightarrow e_{R}^{\mp}W_{R}^{\pm}$ and $e_{R}^{\pm}e_{R}^{\pm}\rightarrow W_{R}^{\pm}W_{R}^{\pm}$ mediated via doubly charged right-handed Higgs triplet scalars will be very rapid in washing out the lepton asymmetry till the temperature drops below the mass of $W_{R}$. At a temperature below the $W_{R}$ mass scale the latter process becomes doubly phase space suppressed. However, in spite of being singly Boltzmann suppressed, the former process stays in equilibrium till a temperature near the electroweak phase transition temperature for $W_{R}$ mass in the $\tev$ range and continues to wash out lepton asymmetry \cite{Dhuria:inprep}. Then the lower limit on $M_{W_{R}}$ for successful resonant leptogenesis will go up beyond the CMS excess range.

Below we consider extensions of the Standard Model motivated by the superstring inspired $E_{6}$ model to explore if the CMS excess signals can be compatible with high-scale leptogenesis.
\section{$E_{6}$-subgroups involving heavy right-handed gauge bosons}
\label{E6}
In this section we explore three effective low energy subgroups of the superstring inspired $E_{6}$ model which involve additional exotic fermions leading to a rich gauge boson phenomenology.  We have already discussed the possibility of producing both the $eejj$ and $e{\slashed p_T jj}$ signals and having sucessful high-scale leptogenesis in the context of low energy subgroups of $E_6$ in Ref. \cite{Dhuria:2015hta} by involving supersymmetric particles. In this letter we assume that supersymmetry gets broken at a very high scale and that supersymmetric partners  do not play any role  in the following analysis.

Under one of the maximal subgroups of $E_{6}$ given by $SU(3)_{C}\times SU(3)_{L} \times SU(3)_{R}$, the fundamental $27$ representation reduces to
\be{\label{4.0.0}}
27= (3, 3, 1)+(3^{\ast}, 1, 3^{\ast})+(1, 3^{\ast} ,3)
\ee
 where $(u, d, h): (3, 3, 1)$ and $(h^{c}, d^{c}, u^{c}): (3^{\ast}, 1, 3^{\ast})$ and $(1, 3^{\ast} ,3)$ corresponds to the leptons. The exotic quark $h$ carries a charge $-\frac{1}{3}$. The other exotic particles are the charge conjugate of $h$, a right-handed neutrino $N^{c}$, two lepton isodoublets $(\nu_{E}, E)$, $(E^{c},N_{E}^{c})$ and $n$. The assignment of the first family is given by
\be{\label{4.0}}
\bpm u \\ d \\ h \epm + \bpm u^{c} & d^{c} & h^{c}\epm +\bpm E^{c} & \nu & \nu_{E} \\ N^{c}_{E} & e & E \\ e^{c} & N^{c} & n \epm,
\ee
where $SU(3)_{L}$ operates vertically and $SU(3)_{R}$ operates horizontally. The $SU(3)_{R,L}$ further decompose to $SU(2) \times U(1)$. There are three different choices for the decomposition of $SU(3)_{R}$ corresponding to three directions of symmetry breaking, which are the familiar $T, U, V$ isospins of $SU(3)$. Below we use the subscript $(R)$ to correspond to these three choices of breaking. These three choices result in three different kinds of heavy right-handed gauge bosons.

\subsection{Case 1.} The $SU(2)_{R}$ doublet is $(d^{c}, u^{c})$ as in the LRSM and $Q=T_{3L}+\frac{1}{2} Y_{L}+T_{3R}+\frac{1}{2} Y_{R}$. Note that $Y_{L}+Y_{R}=(B-L)/2$ holds for all the SM particles and one can extend this as a definition for the new fermions belonging to the fundamental representation of $E_{6}$ to have invariant Yukawa interactions with the SM particles which ensures that all Yukawa and gauge interactions conserve $B-L$.The transformation of the fields under the subgroup $G=SU(3)_c\times SU(2)_L\times SU(2)_R\times U(1)_{B-L}$ is given by
\bea {\label{4.1}}
(u, d)_{L} : (3, 2, 1, \frac{1}{6}),\; (d^{c}, u^{c})_{L} : (\bar{3}, 1, 2, -\frac{1}{6}),\nonumber\\
(\nu_{e}, e)_{L} : (1, 2, 1, -\frac{1}{2}),\; (e^{c}, N^{c})_{L} : (1, 1, 2, \frac{1}{2}),\nonumber\\
h_{L} : (3, 1, 1, -\frac{1}{3}), \; h^{c}_{L} : (\bar{3}, 1, 1, \frac{1}{3}),\nonumber\\
\bpm \nu_{E} & E^{c} \\ E & N^{c}_{E}\epm_{L} : (1, 2, 2, 0),\; n_{L} : (1, 1, 1, 0).
\eea
 If $\nu_{e}$ combines with $N^{c}$ to form the Dirac neutrino then the mass of the $W_{R}^{\pm}$ gets constrained from polarized $\mu^{+}$ decay \cite{Bueno:2011fq}. There will also be a charged current mixing matrix for the quarks in the right-handed sector. Using a form similar to the Kobayashi- Maskawa matrix the $K_{L}- K_{S}$ mass difference can constrain the $W_{R}^{\pm}$ mass\cite{Beall:1981ze, Maiezza:2010ic,  Zhang:2007da}. In Ref. \cite{Senjanovic:2014pva} it was pointed out that a calculation of the mixing matrix for the right-handed quark sector shows that the difference between left and right mixing angles is very small. Kaon decay and neutron electric dipole moment can also give further constraints on the $W_{R}$ mass \cite{Zhang:2007da, Ecker:1983dj}. We have already discussed some of the phenomenological details of the $W_{R}$ decay in connection with the LRSM. Those hold good for this scenario, however one can have more complicated decay modes of $W_{R}$ in the presence of the new exotic fermions. 
 
With the assignment given in Eq. (\ref{4.1}), among the five neutral fermions only $\nu_{e}$ and $N^{c}$ carry nonzero $B-L$. Thus the only source of $B-L$ violation is the Majorana mass of $N^{c}$ which also ensures the small neutrino masses. In order to have successful leptogenesis the decay rate of the Majorana neutrino $N$ must satisfy the out-of-equilibrium condition, namely,
\be{\label{4.1.1}}
\Gamma_{N}< H(T=m_{N}).
\ee
	This translates into the condition that the Majorana  mass of $N_R$ must be many orders of magnitude greater than the $\tev$ scale. On the other hand, the quantum number assignments of $N^{c}$ as given in Eq. (\ref{4.1}) imply that it transforms at low energies. This can result in lepton-number violating interactions involving $W_R$. The associated lepton-number violating scattering processes can wash out the asymmetry produced by leptogenesis at high scale. Therefore successful leptogenesis can not be obtained in this conventional left-right model. Thus we focus below on the two variants where the $SU(2)_{R}$ breaking scale can be much lower ($\sim\tev$ range) independent of the $U(1)_{B-L}$ breaking scale.

\subsection{Case 2.} Another choice for the $SU(2)_{(R)}$ doublet is $(h^{c}, u^{c})$, first pointed out in Ref. \cite{Ma:1986we}. The relevant charge equation is given by $Q=T_{3L}+\frac{1}{2} Y_{L}+T^{\prime}_{3R}+\frac{1}{2} Y^{\prime}_{R}$, where 

\be {\label{4.2}}
T^{\prime}_{3R}=\frac{1}{2} T_{3R}+\frac{3}{2} Y_{R}, \ \ Y^{\prime}_{R}=\frac{1}{2} T_{3R}-\frac{1}{2} Y_{R},
\ee
and we have $T^{\prime}_{3R}+Y^{\prime}_{R}=T_{3R}+Y_{R}$. Note that for interactions involving only the Standard Model particles and gauge bosons (left-handed) the schemes of Case 1 and Case 2 are indistinguishable. In this scenario, often referred to as the Alternative Left Right Symmetric Model (ALRSM) in the literature, the assignments of fields transforming  under the subgroup $G=SU(3)_c\times SU(2)_L\times SU(2)_{R'}\times U(1)_{Y'}$ are given as
\bea {\label{4.3}}
(u, d)_{L} : (3, 2, 1, \frac{1}{6}),\; (h^{c}, u^{c})_{L} : (\bar{3}, 1, 2, -\frac{1}{6}),\nonumber\\
(\nu_{E}, E)_{L} : (1, 2, 1, -\frac{1}{2}),\; (e^{c}, n)_{L} : (1, 1, 2, \frac{1}{2}),\nonumber\\
h_{L} : (3, 1, 1, -\frac{1}{3}),\; d^{c}_{L} : (\bar{3}, 1, 1, \frac{1}{3}),\nonumber\\
\bpm \nu_{e} & E^{c} \\ e & N^{c}_{E}\epm_{L} : (1, 2, 2, 0),\; N^{c}_{L} : (1, 1, 1, 0) ,
\eea
and $Y'=Y_L+Y_R^\prime$. Here also $\nu_{e}$ can pair off with $N^{c}$ to form a Dirac neutrino, but now $N^{c}$ has a trivial transformation under $SU(2)_{R'}$ thus allowing high-scale leptogenesis. Two different assignments for $N^{c}$ are possible determining whether $\nu_{e}$ is massless or massive. For the case where $N^{c}$ has the assignments $B=0$, $L=0$  an exactly massless $\nu_{e}$ is possible, while in the other case $N^{c}$ is assigned $B=0$, $L=-1$ leading to a tiny mass of $\nu_{e}$ via the seesaw mechanism. In this scenario, $e$ is coupled to $n$ via the right-handed charged current, but $n$ being presumably much heavier than the electron, polarized $\mu^{+}$ decay cannot constrain the mass of $W^{ \pm}_{R^{\prime}}$ in contrast to Case 1. Furthermore $W^{ \pm}_{R^{\prime}}$ does not couple to $d$ and $s$ quarks. Consequently, there is no constraint on the mass of  $W^{ \pm}_{R^{\prime}}$ from the $K_{L}- K_{S}$ mass difference in this case. So this model can allow a much lighter $W_{R^\prime}^{\pm}$ as compared to  Case 1. However in this model $D^{0}- \bar{D}^{0}$ mixing can be induced through the $W_{R^{\prime}}$ coupling of the $c$ and $u$ quarks to the  exotic leptoquark $h$ \cite{Ma:1987ji}. The relevant box diagrams are shown in Fig. \ref{fig:wralrm2}.  The amplitude of this mixing induced by these exotic box diagrams can give constraint on the $SU(2)_{R'}$ breaking scale in this model.
 
 \begin{figure}[h]
 	\includegraphics[width=3.4in, height=1.8in]{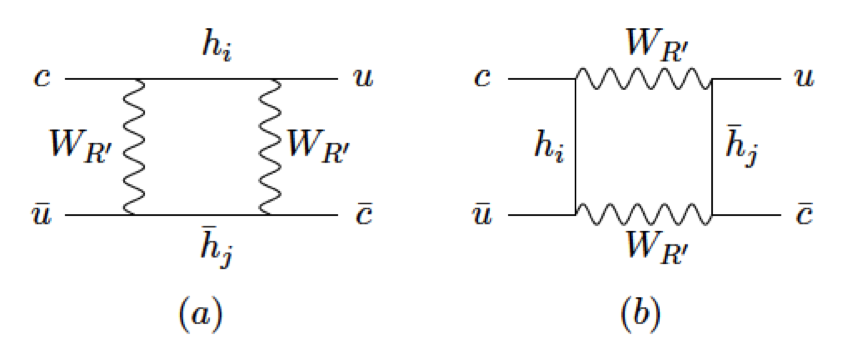}
 	\caption{Box diagrams that can contribute to $D^{0}-{\bar D}^0$ mixing.}
 	\label{fig:wralrm2} 
 \end{figure}
 The interesting point to note here is that in contrast to Case 1 where all the gauge bosons have the assignments $B=0$ and $L=0$, in this case $W^{ -}_{R^{\prime}}$ carries a leptonic charge $L=1$. In this model the coupling of the $W_{R^{\prime}}$ to the fermions reads
\be {\label{4.4}}
\cL=\frac{1}{\sqrt{2}}  g_{R} W^{\mu}_{R^{\prime}} (\bar{h}^{c}\gamma_{\mu}u^{c}_{L} +\bar{E}^{c}\gamma_{\mu}\nu_{L}+\bar{e}^{c}\gamma_{\mu}n_{L}+\bar{N}^{c}_{E}\gamma_{\mu}e_{L})+ \rm{h.c.}
\ee
So the $W_{R^{\prime}}$ is coupled to $h^{c}_{L}$ and $n$ field, in contrast to the coupling with the $d^{c}_{L}$ and $N^{c}$ in the conventional LRSM.

The quantum numbers of $W_{R^{\prime}}$ imply that the usual $u\bar{d}$ scattering in hadronic colliders can not produce $W_{R^{\prime}}$. Furthermore $2M_{W_{R}}> M_{Z^{\prime}}$ forbids the pair production of $W_{R^{\prime}}$ via the decay of the heavy $Z^{\prime}$. The process which can yield a large cross section for $W_{R^{\prime}}$ production is the associated production of $W_{R^{\prime}}$ and leptoquark $h$ via the process $g+u\rightarrow h+W_{R^{\prime}}^{+}$  \cite{Gunion:1987xi}. The relevant diagrams are shown in Fig. \ref{fig:wralrm1}. 
\begin{figure}[h]
	\includegraphics[width=3.4in, height=1.8in]{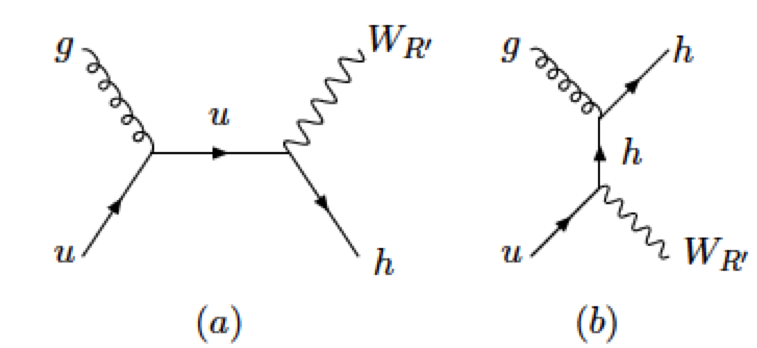}
	\caption{$s$- and $t$-channel Feynman diagrams for $g+u \rightarrow  h+W_{R^{\prime}}$.}
	\label{fig:wralrm1} 
\end{figure}
 The decay modes of the $W_{R^{\prime}}$ can be obtained by using  Eq. (\ref{4.4}).
 \be {\label{4.5}}
 W_{R^{\prime}}\rightarrow   \bar{h}^{c}u^{c}, \; \bar{E}^{c}\nu, \; \bar{e}^{c} n_{L},\; \bar{N}^{c}_{E}e.
  \ee
To keep our discussion fairly general and model independent we only consider the decay modes (of the new exotic particles) mediated by light and heavy gauge bosons (and ignore the decay modes involving Higgs couplings). Examining all the further decay channels of the exotic particles coming from the decay modes of $W_{R^{\prime}}$ listed above immediately shows that the $W_{R^{\prime}}$ decay can not give rise to the $ee+2j$ signal in contrast to Case 1. Thus, this scenario has an appealing feature of allowing high-scale leptogenesis. However, a two electron and two jet signal can not be produced from $W_{R^{\prime}}$ decay.
\subsection{Case 3.} A third way of selecting the $SU(2)_{(R)}$ doublet is $(h^{c}, d^{c})$ \cite{London:1986dk} and the relevant charge equation is given by $Q=T_{3L}+\frac{1}{2} Y_{L}+\frac{1}{2} Y_{N}$, where the $SU(2)_{(R)}$ does not contribute to the electric charge equation and  we will represent it by $SU(2)_{N}$. Once the $SU(2)_{N}$ gets broken, the gauge bosons $W^{\pm}_{N}$ and $Z_{N}$ become massive. The superscript $\pm$ corresponds to the $SU(2)_{N}$ charge. The fields transform under the subgroup $G=SU(3)_c\times SU(2)_L\times SU(2)_N\times U(1)_Y$ as 
\bea {\label{5.1}}
(u, d)_{L} : (3, 2, 1, \frac{1}{6}),\; (h^{c}, d^{c})_{L} : (\bar{3}, 1, 2, \frac{1}{3}),\nonumber\\
(E^{c}, N^{c}_{E})_{L} : (1, 2, 1, \frac{1}{2}),\;  (N^{c}, n)_{L} : (1, 1, 2, 0),\nonumber\\
h_{L} : (3, 1, 1, -\frac{1}{3}),\; u^{c}_{L} : (\bar{3}, 1, 1, -\frac{2}{3}),\nonumber\\
\bpm \nu_{e} & \nu_{E} \\ e & E\epm_{L} : (1, 2, 2, -\frac{1}{2}),\; e^{c}_{L} : (1, 1, 1, 1).
\eea
Similar to case 2, in this scenario also $W_{N}$ has nonzero leptonic charge and zero baryonic charge. Note that in this case $W_{N}$ and $Z_{N}$ can induce $K^{0}- \bar{K}^{0}$ mixing. Mixing between six quarks (three generations) forming $SU(2)_{N}$ doublets 

\be {\label{5.4}}
\bpm \bar{h_{1}} \\ \bar{d} \epm \ \ \bpm \bar{h_{2}} \\ \bar{s} \epm \\ \ \  \bpm \bar{h_{3}} \\ \bar{b} \epm
\ee
\begin{figure}[h]
	\includegraphics[width=3.4in, height=1.8in]{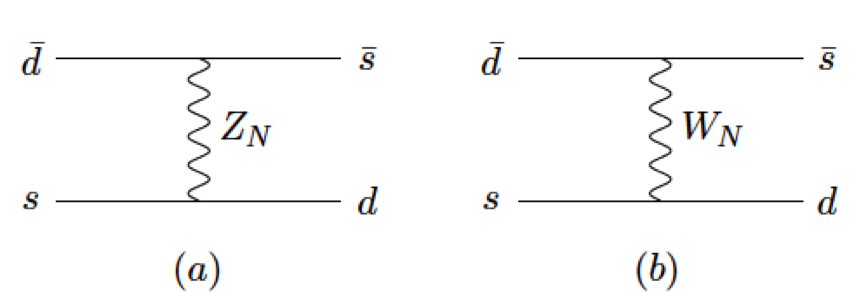}
	\caption{Tree level flavor changing neutral current processes in presence of mixing between six quarks ($d, s, b$ and $h_{i}, i =1,2, 3$).}
	\label{fig:wn3} 
\end{figure}
\begin{figure}[h]
	\includegraphics[width=3.4in, height=1.8in]{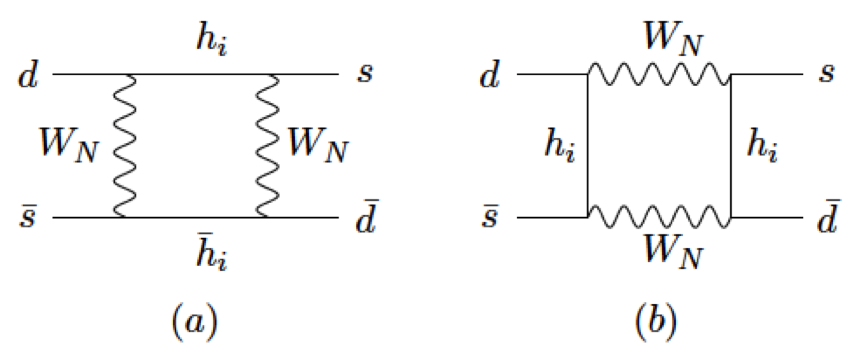}
	\caption{Box diagrams contributing to $\bar{d}s- \bar{s}d$ mixing if only exotic $h_{i}, i=1,2,3$ mix, and $\bar{s}_{L}$ and $\bar{d}_{L}$ have same $T_{3N}$ eigenvalues.}
	\label{fig:wn4} 
\end{figure}
can lead to the tree level Flavor Changing Neutral Current  (FCNC) processes shown in Fig. \ref{fig:wn3} and in such a scenario one can get constraints on the $W_{N}$ mass from the $K_{L}-K_{S}$ mass difference \cite{London:1986dk}. In the absence of  mixing of $\bar{d}$ and $\bar{s}$ with exotic $\bar{h}_{i}$, one can still have a tree level contribution to the kaon mixing. If opposite $T_{3N}$ quantum numbers are assigned to $\bar{d}_{L}$ and $\bar{s}_{L}$ and if they mix then the diagrams shown in Fig. \ref{fig:wn3} are still possible \cite{London:1986dk}. On the other hand if only the exotic $\bar{h}_{i}$ mix and we assign same $T_{3N}$ to $\bar{d}_{L}$ and $\bar{s}_{L}$ then the box diagrams shown in Fig. \ref{fig:wn4} result \cite{London:1986dk}.  Likewise in the leptonic sector considering $SU(2)_{N}$ doublets 
\be {\label{5.5}}
\bpm E \\ e \epm \ \  \bpm M \\ \mu \epm \ \ \bpm T \\ \tau \epm,
\ee
even if mixing between the ordinary and exotic fermions is absent, the process $\mu\rightarrow e \gamma$ can be possible if mixing between the exotic fermions is present \cite{London:1986dk} as shown in Fig. \ref{fig:wn5}. 

\begin{figure}[h]
	\includegraphics[width=3.1in, height=1.5in]{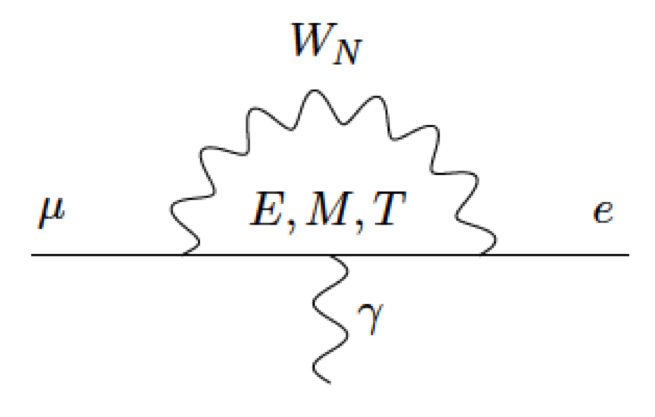}
	\caption{Loop diagrams involving exotic fermions (mixing among themselves) and $W_{N}$ leading to $\mu \rightarrow e \gamma$.}
	\label{fig:wn5} 
\end{figure}
The coupling of the $W_{N}$ to the fermions reads
\be {\label{5.2}}
\cL=\frac{1}{\sqrt{2}}  g_{R} W^{\mu}_{N} (\bar{h}\gamma_{\mu}d_{R} +\bar{e}\gamma_{\mu} E_{L}+\bar{\nu}\gamma_{\mu}(\nu_{E})_{L}+\bar{N}^{c} \gamma_{\mu}n_{L})+ \rm{h.c.}
\ee
\begin{figure}[h]
	\includegraphics[width=3in,height=1.5in]{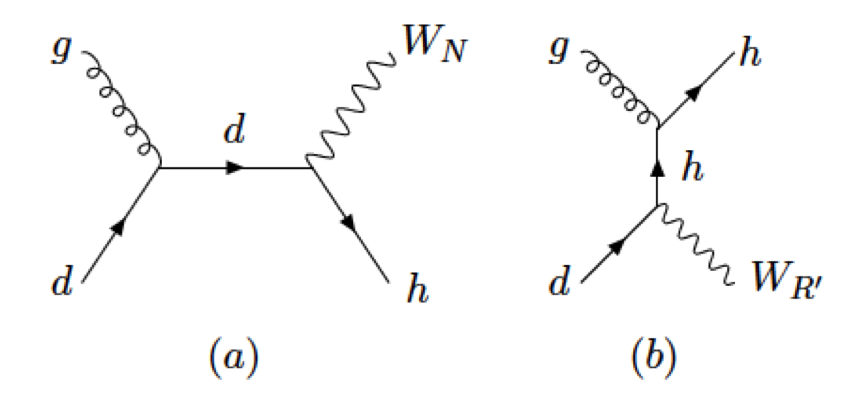}
	\caption{$s$- and $t$-channel Feynman diagrams for $g+d \rightarrow  h+W_{N}$.}
	\label{fig:wn1} 
\end{figure}
Following similar arguments as in Case 2, one can not produce $W_{N}$ via the usual Drell-Yan mechanism or from the decay of heavy $Z_{N}$. The process $g+d\rightarrow h+W_{N}$ can yield a large cross section for $W_{N}$ production via the diagrams shown in Fig. \ref{fig:wn1} \cite{Rizzo:1988bv}. Pair production of $W_{N}$ can take place via the process $e^{+}e^{-}\rightarrow W^{+}_{N}W^{-}_{N}$ \cite{Rizzo:1988bv}. The relevant diagrams are shown in Fig. \ref{fig:wn2}.  This process is particularly sensitive to the underlying  gauge structure and cancellations between the given amplitudes. Thus it can serve as a probe for the non-abelian $SU(2)_{N}$ gauge theory.

\begin{figure}[h]
	\includegraphics[width=3.2in,height=1.4in]{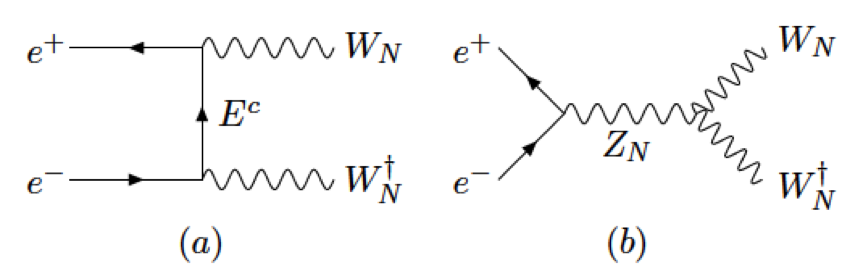}
	\caption{$t-$ and $s-$channel Feynman diagrams for $e^{+}e^{-} \rightarrow W^{+}_{N}W^{-}_{N}$.}
	\label{fig:wn2} 
\end{figure}
The decay modes of the $W_{N}$ can be obtained from Eq. (\ref{5.2}) as
\be {\label{5.3}}
W_{N}\rightarrow   \bar{h}d, \; \bar{e}E, \; \bar{\nu} \nu_{E},\; \bar{N}^{c}n_{L}.
\ee 
Like in Case 2, an inspection of all the further decays of the exotic particles for the decay modes of $W_{N}$ listed above tells us that an  $ee+2j$ signal can not be obtained from the decay of $W_{N}$. Moreover from the assignments of Eq. (\ref{5.1}) it follows that $N^{c}$ transforms as a doublet under $SU(2)_{N}$ and hence for low-energy $SU(2)_{N}$ breaking, following the same logic as in Case 1, the possibility of successful leptogenesis is ruled out. 
\section{Conclusions}
\label{conc}
 We have  reviewed the non-supersymmetric versions of  the (Extended) Left-Right Symmetric Model and the models appearing as the low-energy subgroups of the superstring motivated $E_{6}$ group which can have 
 low-scale $SU(2)_{(R)}$ breaking. Our aim was to examine if a signal like the CMS $eejj$ excess can be explained from these models while allowing leptogenesis.
 
 In the LRSM and any extension of it with multiple U(1)'s, for hierarchical neutrino masses ($M_{N_{3R}}\gg M_{N_{2R}}\gg M_{N_{1R}}=m_{N}$) the possibility of generating the required baryon asymmetry of the universe from high-scale leptogenesis is ruled out if the $W_{R}$ mass lies in the $\tev$ range as indicated by the CMS events. Recently, it was shown that the required lepton asymmetry can be produced even for  a substantially low value of the $W_R$ mass ($M_{W_{R}}>3 \tev$) \cite{Dev:2014iva} by considering relatively large Yukawa couplings in the context of resonant leptogenesis. However we have mentioned that certain lepton-number violating scattering processes involving the doubly charged Higgs triplet can wash out the lepton asymmetry below the $B-L$ breaking scale till the electroweak phase transition thus ruling out the possibility of resonant leptogenesis for the mass range of $W_R$ as indicated by the CMS excess signal. Therefore we have then considered low energy subgroups of the superstring motivated $E_6$ group involving new exotic fermions and a low-energy $SU(2)_{(R)}$ gauge sector. Amongst all low energy subgroups considered in the analysis there is only one choice of $SU(2)_{(R)}$ which allows high-scale leptogenesis. However, this particular choice cannot account for the excess signal seen at CMS. So this together with our  consideration of high-scale and $\tev$-scale resonant leptogenesis for the LRSM and its extensions implies that a pre-electroweak phase transition leptogenesis scenario can not generate the baryon asymmetry in the non-supersymmetric models under consideration. Thus one needs to look for post-sphaleron mechanisms to explain the observed baryon asymmetry of the universe. To this end, possibilities like neutron-antineutron oscillations can be explored \cite{Phillips:2014fgb}.


\begin{thebibliography}{}
  \bibitem{Pati:1974yy} 
 J.~C.~Pati and A.~Salam,
 Phys.\ Rev.\ D {\bf 10}, 275 (1974)
 [Erratum-ibid.\ D {\bf 11}, 703 (1975)];  R.~N.~Mohapatra and J.~C.~Pati,
 Phys.\ Rev.\ D {\bf 11}, 2558 (1975); G.~Senjanovic and R.~N.~Mohapatra,
 Phys.\ Rev.\ D {\bf 12}, 1502 (1975).

 
 \bibitem{Khachatryan:2014dka} 
 V.~Khachatryan {\it et al.}  [CMS Collaboration],
Eur.Phys.J. C {\bf 74} (2014) 11, 3149
 [arXiv:1407.3683 [hep-ex]].


 
    \bibitem{Deppisch:2014qpa}
    F.~F.~Deppisch, T.~E.~Gonzalo, S.~Patra, N.~Sahu and U.~Sarkar,
    Phys.\ Rev.\ D {\bf 90}, 053014 (2014)
    [arXiv:1407.5384 [hep-ph]], [arXiv:1410.6427 [hep-ph]].
    
  
    \bibitem{Saavedra2014}
    J.~A.~Aguilar-Saavedra and F.~R.~Joaquim,
    Phys.\ Rev.\ D {\bf 90}, 115010 (2014)
    [arXiv:1408.2456 [hep-ph]].

\bibitem{Heikinheimo-2014} 
 M.~Heikinheimo, M.~Raidal and C.~Spethmann,
 Eur.\ Phys.\ J.\ C {\bf 74}, 3107 (2014)
 [arXiv:1407.6908 [hep-ph]].
 
 \bibitem{Fukugita:1986hr} 
 M.~Fukugita and T.~Yanagida,
 Phys.\ Lett.\ B {\bf 174}, 45 (1986);
 for recent reviews see e.g. 
 S.~Davidson, E.~Nardi and Y.~Nir,
 Phys.\ Rept.\  {\bf 466}, 105 (2008)
 [arXiv:0802.2962 [hep-ph]];
 C.~S.~Fong, E.~Nardi and A.~Riotto,
 Adv.\ High Energy Phys.\  {\bf 2012}, 158303 (2012)
 [arXiv:1301.3062 [hep-ph]].
 
 \bibitem{Minkowski:1977sc} 
 P.~Minkowski,
 Phys.\ Lett.\ B {\bf 67}, 421 (1977);
 M.~Gell-Mann, P.~Ramond and R.~Slansky,
 Conf.\ Proc.\ C {\bf 790927}, 315 (1979)
 [arXiv:1306.4669 [hep-th]];
 T.~Yanagida,
 Conf.\ Proc.\ C {\bf 7902131}, 95 (1979);
 R.~N.~Mohapatra and G.~Senjanovic,
 Phys.\ Rev.\ Lett.\  {\bf 44}, 912 (1980);
 R.~N.~Mohapatra and G.~Senjanovic,
 Phys.\ Rev.\ D {\bf 23}, 165 (1981).
 
 \bibitem{Kuzmin:1985mm} 
 V.~A.~Kuzmin, V.~A.~Rubakov and M.~E.~Shaposhnikov,
 Phys.\ Lett.\ B {\bf 155}, 36 (1985).
 
\bibitem{Ma:1998dx} 
E.~Ma and U.~Sarkar,
Phys.\ Rev.\ Lett.\  {\bf 80}, 5716 (1998)
[hep-ph/9802445];
G.~Lazarides and Q.~Shafi,
Phys.\ Rev.\ D {\bf 58}, 071702 (1998)
[hep-ph/9803397];
T.~Hambye, E.~Ma and U.~Sarkar,
Nucl.\ Phys.\ B {\bf 602}, 23 (2001)
[hep-ph/0011192].
 
 \bibitem{Ma:1998sq} 
 E.~Ma, S.~Sarkar and U.~Sarkar,
 Phys.\ Lett.\ B {\bf 458}, 73 (1999)
 [hep-ph/9812276].
 
   \bibitem{Flanz:1994yx} 
   M.~Flanz, E.~A.~Paschos and U.~Sarkar,
   Phys.\ Lett.\ B {\bf 345}, 248 (1995)
   [Erratum-ibid.\ B {\bf 382}, 447 (1996)]
   [hep-ph/9411366];
   M.~Flanz, E.~A.~Paschos, U.~Sarkar and J.~Weiss,
   Phys.\ Lett.\ B {\bf 389}, 693 (1996)
   [hep-ph/9607310];
   A.~Pilaftsis,
   Phys.\ Rev.\ D {\bf 56}, 5431 (1997)
   [hep-ph/9707235];
   E.~Roulet, L.~Covi and F.~Vissani,
   Phys.\ Lett.\ B {\bf 424}, 101 (1998)
   [hep-ph/9712468];
   W.~Buchmuller and M.~Plumacher,
   Phys.\ Lett.\ B {\bf 431}, 354 (1998)
   [hep-ph/9710460];
   M.~Flanz and E.~A.~Paschos,
   Phys.\ Rev.\ D {\bf 58}, 113009 (1998)
   [hep-ph/9805427];
   T.~Hambye, J.~March-Russell and S.~M.~West,
   JHEP {\bf 0407}, 070 (2004)
   [hep-ph/0403183];
   A.~Pilaftsis and T.~E.~J.~Underwood,
   Nucl.\ Phys.\ B {\bf 692}, 303 (2004)
   [hep-ph/0309342].
   
    \bibitem{Frere:2008ct} 
    J.~M.~Frere, T.~Hambye and G.~Vertongen,
    JHEP {\bf 0901}, 051 (2009)
    [arXiv:0806.0841 [hep-ph]].
   
   \bibitem{Dev:2014iva} 
   P.~S.~Bhupal Dev, C.~H.~Lee and R.~N.~Mohapatra,
  Phys.Rev. D {\bf 90} (2014) 9, 095012
   [arXiv:1408.2820 [hep-ph]].
   
    \bibitem{Dhuria:inprep} 
    M.~Dhuria, C.~Hati, R.~Rangarajan and U.~Sarkar,
    Manuscript in preparation.
   
   \bibitem{Dhuria:2015hta} 
   M.~Dhuria, C.~Hati, R.~Rangarajan and U.~Sarkar,
    to appear in Phys.\ Rev.\ D (2015)
   [arXiv:1501.04815 [hep-ph]].

\bibitem{Bueno:2011fq} 
J.~F.~Bueno {\it et al.}  [TWIST Collaboration],
Phys.\ Rev.\ D {\bf 84}, 032005 (2011)
[arXiv:1104.3632 [hep-ex]].

\bibitem{Beall:1981ze} 
G.~Beall, M.~Bander and A.~Soni,
Phys.\ Rev.\ Lett.\  {\bf 48}, 848 (1982).

\bibitem{Maiezza:2010ic} 
A.~Maiezza, M.~Nemevsek, F.~Nesti and G.~Senjanovic,
Phys.\ Rev.\ D {\bf 82}, 055022 (2010)
[arXiv:1005.5160 [hep-ph]].

\bibitem{Zhang:2007da} 
Y.~Zhang, H.~An, X.~Ji and R.~N.~Mohapatra,
Nucl.\ Phys.\ B {\bf 802}, 247 (2008)
[arXiv:0712.4218 [hep-ph]].

\bibitem{Senjanovic:2014pva} 
  G.~Senjanović and V.~Tello,
  Phys.\ Rev.\ Lett.\  {\bf 114}, 071801 (2015)
  [arXiv:1408.3835 [hep-ph]].

\bibitem{Ecker:1983dj} 
G.~Ecker, W.~Grimus and H.~Neufeld,
Nucl.\ Phys.\ B {\bf 229}, 421 (1983).
 
 \bibitem{Ma:1986we} 
 E.~Ma,
 Phys.\ Rev.\ D {\bf 36}, 274 (1987).
  
  \bibitem{Ma:1987ji} 
  E.~Ma,
  Mod.\ Phys.\ Lett.\ A {\bf 3}, 319 (1988).
 
 \bibitem{Gunion:1987xi} 
 J.~F.~Gunion, J.~L.~Hewett, E.~Ma, T.~G.~Rizzo, V.~D.~Barger, N.~Deshpande and K.~Whisnant,
 Int.\ J.\ Mod.\ Phys.\ A {\bf 2}, 1199 (1987).

 \bibitem{London:1986dk} 
 D.~London and J.~L.~Rosner,
 Phys.\ Rev.\ D {\bf 34}, 1530 (1986).
  
  \bibitem{Rizzo:1988bv} 
  T.~G.~Rizzo,
  Phys.\ Rev.\ D {\bf 38}, 71 (1988)
  [Addendum-ibid.\ D {\bf 39}, 3490 (1989)].
  
\bibitem{Phillips:2014fgb} 
  D.~G.~Phillips, II, W.~M.~Snow, K.~Babu, S.~Banerjee, D.~V.~Baxter, Z.~Berezhiani, M.~Bergevin and S.~Bhattacharya {\it et al.},
  [arXiv:1410.1100 [hep-ex]].


	
\end{thebibliography}
\end{document}